\documentclass[prb,letterpaper,aps,floatfix,twocolumn,superscriptaddress]{revtex4-1}
\usepackage{graphicx}
\usepackage{amsmath}
\newcommand{\beq}{\begin{equation}}
\newcommand{\eeq}{\end{equation}}

\newcommand{\br}{{{\bf{r}}}}

\newcommand{\bA}{{\bf{A}}}
\newcommand{\bB}{{\bf{B}}}

\newcommand{\beqa}{\begin{eqnarray}}
\newcommand{\eeqa}{\end{eqnarray}}

\newcommand{\ua}{\uparrow}
\newcommand{\da}{\downarrow}

\newcommand{\dg}{{\dag}}
\newcommand{\pdg}{{\vphantom\dag}}
\newcommand{\btau}{{\boldsymbol \tau}}
\newcommand{\bsigma}{{\boldsymbol \sigma}}
\newcommand{\bpi}{{\boldsymbol \pi}}
\newcommand{\bnabla}{{\boldsymbol \nabla}}
\begin{document}
\title{Thin topological insulator film in a perpendicular magnetic field}
\author{A.A. Zyuzin}
\affiliation{Department of Physics and Astronomy, University of Waterloo, Waterloo, Ontario 
N2L 3G1, Canada} 
\author{A.A. Burkov}
\affiliation{Department of Physics and Astronomy, University of Waterloo, Waterloo, Ontario N2L 3G1, Canada}
\affiliation{Department of Physics, California Institute of Technology, Pasadena, California 91125, USA}
\date{\today}
\begin{abstract}
We report on a study of an ultrathin topological insulator film with hybridization between the top and bottom 
surfaces, placed in a quantizing perpendicular magnetic field. We calculate the full Landau level spectrum of the 
film as a function of the applied magnetic field and the magnitude of the hybridization matrix element, 
taking into account both the orbital and the Zeeman spin splitting effects of the field.
For an undoped film, we find a quantum phase transition between a state with a zero Hall conductivity 
and a state with a quantized Hall conductivity equal to $e^2/h$, as a function of the magnitude of the 
applied field. The transition is driven by the competition between the Zeeman and the hybridization energies. 
\end{abstract}
\maketitle
\section{Introduction}
\label{sec:intro}
Topological insulator (TI) is a new phase of matter, which has recently been discovered in materials, 
long known and extensively studied in the fields of traditional semiconductor physics (HgTe) and the physics of 
thermoelectrics ($\textrm{Bi}_2 \textrm{Se}_3$ and $\textrm{Bi}_2 \textrm{Te}_3$).~\cite{Kane05,Hasan09,Kane10} 
Somewhat uncharacteristically of the ``standard" order of things in condensed matter physics, TI phase was first predicted 
to occur in these materials theoretically~\cite{Kane05} and only later seen experimentally.~\cite{Hasan09}
One obvious reason for this is that the main experimental manifestation of the TI phase, namely the occurence of metallic edge 
states in insulating samples, is a somewhat subtle effect, unlike, e.g., the spectacular quantization of the Hall conductivity 
in a close relative of TI, the two-dimensional electron gas  (2DEG) in the quantum Hall effect (QHE) regime. 
While theoretically TI phase can in fact be characterized by a quantized physical quantity, 
the topological magnetoelectric susceptibility,~\cite{Qi08,Essin09} measuring this quantity experimentally is not straightforward.
Currently, the only realistic proposal involves measurement of the magneto-optical response in a thin TI film when the time-reversal 
symmetry is broken, with the Faraday and Kerr rotation angles predicted to acquire universal values in this 
case.~\cite{Tse10-1,Tse10-2,Maciejko10} 

As far as transport phenomena in TI are concerned, theoretical effort has mostly been concentrated on studying the effects of the 
characteristic spin-momentum locking of the helical TI surface states, which has an obvious potential in 
spintronics.~\cite{Raghu10,Garate10,Garate10-1,Sengupta10-1,Sengupta10-2,Nagaosa10,Nagaosa10-1,Burkov10,DasSarma10,Nagaosa10-2}
An important practical issue, relevant to all transport studies of TI, is that all currently known TI materials, while theoretically
insulators, in reality are metallic in the bulk. This happens due to unavoidable doping, introduced by impurities and crystal lattice defects. 
One possible way to deal with problem is to grow TI samples in the form of thin films.~\cite{ZhangG09,Peng09,He10,Sudbo09}
This allows to both directly reduce the bulk contribution to transport by simply reducing 
the bulk volume and to control the carrier concentration in the bulk through gating.~\cite{Steinberg10,Checkelsky10,Chen10}

Ultrathin TI films also have interesting physical properties, such as the already mentioned above universal magneto-optical 
response,~\cite{Tse10-1,Tse10-2,Maciejko10} possible excitonic superfluidity,~\cite{Seradjeh09} strongly 
improved thermoelectric performance,~\cite{Ghaemi10} quantum spin Hall~\cite{Lu10,Liu10} and quantum anomalous Hall (QAH)
effects,~\cite{Yu10} which have so far been explored only theoretically.
Of particular interest are the effects which arise in the ultrathin limit, when the top and bottom surfaces of a TI film start to 
hybridize.~\cite{He10}
We have recently pointed out~\cite{Zyuzin10} that magnetic response, in particular response to field, applied in the plane of the film, is highly 
nontrivial in this limit, with the film undergoing a topological insulator to semimetal quantum phase transition as a function of the field. 

In this paper we study the properties of TI thin films in a {\em perpendicular} quantizing magnetic field.
We calculate the full Landau level (LL) spectrum of a TI film with hybridization between the top and bottom surfaces and the Hall conductivity.
One would normally expect the Hall conductivity to be zero in an undoped charge-neutral film, which is indeed what happens when hybridization is stronger than the Zeeman splitting.  
As the magnetic field is increased, however, we find a quantum phase transition, at which the Hall conductivity jumps from zero to 
$e^2/h$. The transition happens when the Zeeman energy becomes larger than the hybridization energy and is shown here to be a direct consequence of the well-known characteristic feature of the Dirac-fermion LL spectrum, namely the existence a zero-energy LL.
The transition we find is analogous to the transition to the QAH state discussed in Ref.[\onlinecite{Yu10}] in the context of 
TI films, doped with magnetic impurities. Here we demonstrate that the same physics can be realized in a more straightforward
and currently experimentally accessible way, simply utilizing an applied magnetic field.  
Indeed, experimental work on magnetotransport properties of thin TI films in quantizing perpendicular magnetic  field has already appeared in 
the literature.~\cite{Analytis10,Ando10,Ong11,Sacepe11} 
\section{Landau level spectrum of an ultrathin TI film}
\label{sec:2}
We start from the following Hamiltonian of an ultrathin TI film in the presence of a perpendicular magnetic field:
\beqa
\label{eq:1}
{\cal H}&=&\int d \br \, \Psi^\dag(\br)  \left[ v_F \tau^z \left(\hat z \times \bsigma \right) \cdot \left(-i \bnabla + \frac{e}{c} \bA \right) + 
\Delta_z \sigma^z \right. \nonumber \\
&+&\left. \Delta_t \tau^x \right] \Psi^\pdg(\br). 
\eeqa
Here $v_F$ is the surface Dirac cone Fermi velocity, $\Delta_z = g \mu_B B /2$ is the Zeeman energy, associated 
with the applied magnetic field $\bB = B \hat z$ (we will assume $B \geq 0$ henceforth), $\Delta_t$ is the hybridization matrix element 
(we will also assume $\Delta_t \geq 0$)
and $\hbar =1$ units are used. 
We have introduced Pauli matrices $\bsigma$ and $\btau$ to describe the real spin and the {\em which surface} pseudospin
degrees of freedom and suppressed the explicit spin and pseudospin indices for clarity of notation. 

To diagonalize Eq.(\ref{eq:1}), we choose Landau gauge for the vector potential $\bA = x B \hat y$ and define LL
ladder operators in the standard way as:
\beq
\label{eq;1a}
a = (\ell/\sqrt{2}) (\pi_x - i \pi_y), \,\,  a^\dg = (\ell/\sqrt{2}) (\pi_x + i \pi_y),
\eeq
where $\ell = \sqrt{c/ e B}$ is the magnetic length and $\bpi = -i \bnabla+ (e/c) \bA$ is the kinetic momentum. 
The single-particle Hamiltonian operator can then be written in terms of the ladder operators as:
\beq
\label{eq:2}
H = \frac{i \omega_B}{\sqrt{2}} \tau^z (\sigma^+ a - \sigma^- a^\dag) + \Delta_z \sigma^z + \Delta_t \tau^x,
\eeq
where we have introduced a characteristic frequency $\omega_B = v_F / \ell$, which plays a role,
analogous to the cyclotron frequency in the LL spectrum of a regular 2DEG.

It is clear from Eq.(\ref{eq:2}) that the single-particle eigenstates have the following general form:
\beqa
\label{eq:3}
| n \alpha  s \rangle&=&u^{\alpha  s}_{n T \ua} | n-1, T, \ua \rangle + u^{\alpha s}_{n T \da} | n, T, \da \rangle  \nonumber \\
&+& u^{\alpha  s}_{n B \ua} | n-1, B, \ua \rangle + u^{\alpha s}_{n B \da} | n, B, \da \rangle.
\eeqa
Here $| n, T(B), \ua(\da) \rangle$ is the $n$-th LL eigenstate on the top (bottom) surface with spin up (down),
$\alpha=0, 1$ and $s = \pm$ label the four eigenstates of Eq.(\ref{eq:2}), corresponding to each LL index $n = 0,\ldots,\infty$, and 
$u^{\alpha  s}_n$ are the corresponding complex four-component spinor wavefunctions.  
We have also suppressed the intra-LL orbital label for brevity.
The physical meaning of the $\alpha$ and $s$ indices will become clear shortly. 

The problem of finding the eigenstates of Eq.(\ref{eq:2}) thus reduces to diagonalizing the following $4\times 4$ matrix:
\beqa
\label{eq:4}
\left(
\begin{array}{cccc}
\Delta_z & - i \omega_B \sqrt{2 n} & \Delta_t & 0 \\
i \omega_B \sqrt{2 n}  & - \Delta_z &  0 & \Delta_t \\
\Delta_t & 0 & \Delta_z &  i \omega_B \sqrt{2 n} \\
0 & \Delta_t &  -i \omega_B \sqrt{2 n} & - \Delta_z
\end{array}
\right).
\eeqa
Diagonalizing (\ref{eq:4}), we find the following LL spectrum:
\beq
\label{eq:5}
\epsilon_{n \alpha \pm} = (-1)^{\alpha} \sqrt{2 \omega_B^2 n + (\Delta_z \pm \Delta_t)^2}.
\eeq
The corresponding spinor wavefunctions are given by:
\beq
\label{eq:6}
u^{\alpha s}_n = \left[ i s (-1)^{\alpha}  f_{n \alpha s +}, -s f_{n \alpha s -}, i (-1)^{\alpha} f_{n \alpha s +}, f_{n \alpha s -} \right],  
\eeq
where
\beq
\label{eq:7}
f_{n \alpha s \pm} = \frac{1}{2} \sqrt{1 \pm \frac{\Delta_z + s \Delta_t}{\epsilon_{n \alpha s}}}.
\eeq
The LL spectrum thus consists of two sets: electron-like ($\epsilon_{n 0 \pm}$) and hole-like 
($\epsilon_{n 1 \pm}$) Landau levels. Within each set, every LL is further split into a doublet ($\pm$). 
Note that this splitting results from the presence of both the Zeeman spin splitting and the hybridization, and vanishes 
if any one of them is zero (except for the $n = 0$ LL, which is split whenever $\Delta_t$ is nonzero). 

The $n = 0$ LL are special. As seen from Eq.~(\ref{eq:3}), electrons in these levels are fully spin polarized, 
i.e. only the spin-down states are occupied. Correspondingly, the $n = 0$ level is only split into two sublevels, unlike all other LL, 
which are split into four. This is the well-known ``zero-mode anomaly" of the Dirac-fermion LL
spectrum,~\cite{Jackiw84,Semenoff84,Haldane88} which will play an important role in our story. 

\begin{figure}[t]
\includegraphics[width=8cm]{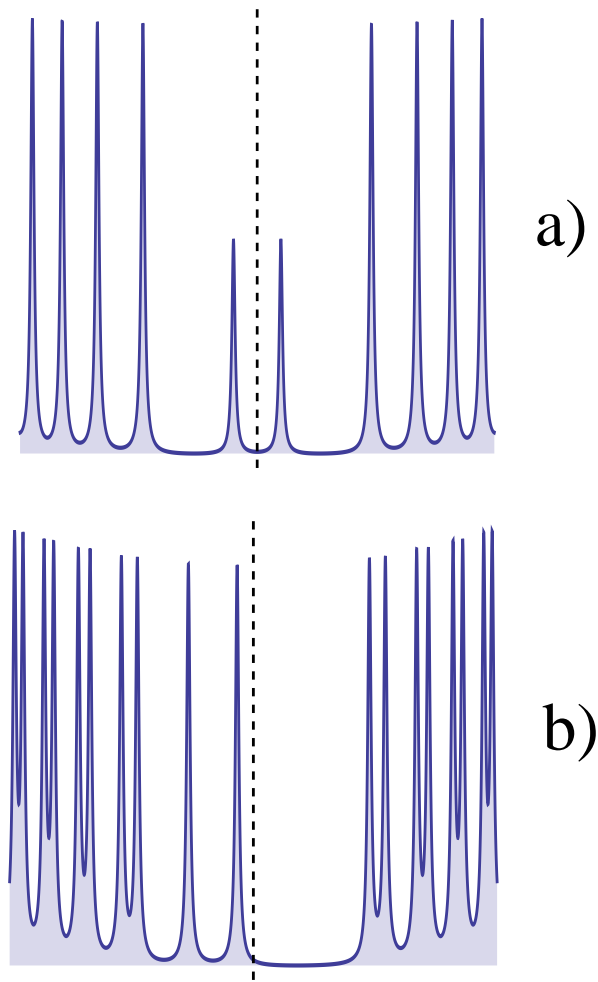}
\caption{(Color online) LL density of states of a thin TI film (broadening added by hand).
Dashed line represents the zero of energy, set at the Dirac point position in the absence of the magnetic field.   
(a) $\Delta_z = 0$, $\Delta_t > 0$ case. Only the $n = 0$ LL is split and the spectrum is particle-hole 
symmetric. 
(b) $\Delta_z > \Delta_t > 0$ case. Both $n = 0$ LL are hole-like and particle-hole symmetry of the spectrum is broken.
All LL splittings in the figure are exaggerated relative to the main LL spacing.} 
\label{fig:1}
\end{figure}

Explicitly, the two $n = 0$ LL wavefunctions are given by:
\beq
\label{eq:8a}
u^{0-}_0 = \frac{1}{\sqrt{2}} (0, 1, 0, 1), \,\,\, u^{1+}_0 = \frac{1}{\sqrt{2}} (0, -1, 0, 1), 
\eeq
when $\Delta_z < \Delta_t$. 
The corresponding LL energies in this case are:
\beq
\label{eq:8b}
\epsilon_{00-} = |\Delta_z - \Delta_t |,\,\, \epsilon_{01+} = - |\Delta_z + \Delta_t |,
\eeq 
On the other hand, when $\Delta_z > \Delta_t$,  $u^{0-}_0$ is replaced by $u^{1-}_0$:
\beq
\label{eq:8c}
u^{1-}_0 = \frac{1}{\sqrt{2}} (0, 1, 0, 1),
\eeq
with the corresponding energy given by:
\beq
\label{eq:8d}
\epsilon_{01-} = - |\Delta_z - \Delta_t |.
\eeq 
Thus one of the two $n = 0$ sublevels moves from the set of electron-like levels into the set of hole-like levels. 
This change of the character of the LL spectrum as a function of $\Delta_z$ is illustrated in Fig.~\ref{fig:1}. 
As we demonstrate below, the transition to purely hole-like $n = 0$ LL is manifested in a jump of the Hall conductivity at fixed chemical potential $\epsilon_F = 0$ by $e^2/h$.

\section{Hall conductivity}
\label{sec:3}
Given the above LL spectrum, we can now evaluate the Hall conductivity of the TI film. 
Assuming the Fermi level to be always in a gap between the LL, the Kubo-formula expression for
the Hall conductivity reads:
\beqa
\label{eq:9}
\sigma_{xy}&=&\frac{\omega_B^2 e^2}{2 \pi} \sum_{n \alpha s \ne n' \alpha' s'} \textrm{Im} \left[ \langle n  \alpha s | \tau^z \sigma^y |
n'  \alpha'  s' \rangle \right. \nonumber \\
&\times& \left. \langle n'  \alpha'  s' | \tau^z \sigma^x | n  \alpha  s \rangle \right]
\frac{n_F(\epsilon_{n \alpha s}) - n_F(\epsilon_{n' \alpha' s'})}{(\epsilon_{n \alpha s} - \epsilon_{n' \alpha' s'})^2}, \nonumber \\ 
\eeqa
where $n_F$ is the Fermi-Dirac distribution (we will assume zero temperature in all calculations). 
The matrix elements in Eq.(\ref{eq:9}) are nonzero only when $s = s'$ and $n' = n \pm 1$. 
Evaluating the matrix elements, we obtain the following expression for the Hall conductivity:
\beqa
\label{eq:10}
\sigma_{xy}&=& \frac{e^2}{4 \pi}  \sum_{n = 0}^{\infty} \sum_{\alpha =0,1} \sum_{s = \pm} \left\{ (2 n +1) 
\left[n_F(\epsilon_{n \alpha s}) - n_F(\epsilon_{n+1 \alpha s}) \right] \right. \nonumber \\
&+&\left. (\Delta_z + s \Delta_t) \left[ \frac{n_F(\epsilon_{n+1 \alpha s})}{\epsilon_{n+1 \alpha s}} - 
 \frac{n_F(\epsilon_{n \alpha s})}{\epsilon_{n \alpha s}} \right] \right\}. 
 \eeqa 
 The first term in Eq.(\ref{eq:10}) corresponds to the well-known expression for the Dirac-fermion Hall conductivity. 
 Assuming $n$ electron-like LL are filled and neglecting the Zeeman and hybridization splittings, which are negligible for all $n > 0$ LL, 
 this term gives: 
 \beq
 \label{eq:11}
 \sigma_{xy} = \frac{e^2}{h} (2 n+1),\,\, n\geq 1, 
 \eeq
 where we have restored $\hbar$ for clarity. This result is the same (up to a factor of $1/2$, accounting for half the degeneracy) 
 as the well-known result for the Hall conductivity of graphene.~\cite{Gusynin05}

The contribution of the $n =0$ LL is contained in the second term in Eq.(\ref{eq:10}), which also contains 
the physics, unique to TI thin films. 
The effect of this term is most easily seen at the point $\epsilon_F =0$, which corresponds to the charge-neutrality
point in the limit $\Delta_z = 0$ ($\Delta_z > 0$ breaks particle-hole symmetry, as seen from Fig.~\ref{fig:1}). 
In this case we obtain the following expression for the Hall conductivity (note again that $\Delta_z, \Delta_t \geq 0$):
\beq
\label{eq:12}
\sigma_{xy} = \frac{e^2}{2 h} \left[ \textrm{sign}(\Delta_z - \Delta_t) + 1 \right].
\eeq
Thus one can see that $\sigma_{xy}$, evaluated at fixed Fermi energy $\epsilon_F = 0$, 
jumps from $0$ ($\Delta_z  < \Delta_t$) to $\pm e^2/h$ ($\Delta_z > \Delta_t$) as a function of the applied magnetic field. 
This transition is the result of the change of the character of the $n=0$ doublet of LL, which happens 
as a function of $\Delta_z/ \Delta_t$, and which was described in Section~\ref{sec:2}.
To reiterate, when $\Delta_z < \Delta_t$, one of the $n = 0$ LL ($u^{0-}$) is electron-like, while the other one ($u^{1+}$) is hole-like.
When $\Delta_z > \Delta_t$, both $n = 0$ LL become hole-like ($u^{1\pm}$), and the Hall 
conductivity thus increases by $e^2/h$. 
This change in the value of the Hall conductivity is of course only observed when the Fermi level is initially within the hybridization gap 
of the $n = 0$ LL. 

It is important to note that the above transition at a fixed Fermi energy is associated with a small increase of the (two-dimensional) 
electron density by $\delta n_e \sim 1/ 2 \pi \ell^2|_{\Delta_z = \Delta_t} = m \Delta_t / \pi$,
since the high-field side of the transition $\Delta_z > \Delta_t$ differs from the low-field side by an extra filled LL. 
If the charge density is held fixed and one starts from the low-field limit when the Fermi level is in the hybridization gap between 
the two $n = 0$ Landau sublevels, no transition occurs upon increasing the field, as the Fermi level moves with the field and always stays 
in the gap. 
However, the fixed Fermi energy situation might in fact be more physically relevant for a sample in a transport-measurement setup, attached to 
external conducting leads (admittedly, this is a subtle issue and what actually happens may depend on details of the 
experimental setup and various characteristics of the sample). 
 
\section{Discussion and conclusions}
\label{sec:4}
In this paper we have studied the properties of an ultrathin TI film, in which the top and bottom surfaces are hybridized, 
placed in a perpendicular magnetic field. 
We have calculated the LL spectrum of such a film, taking into account the Zeeman spin splitting. 
We have also evaluated the Hall conductivity of the film and demonstrated that the Hall conductivity 
has a nontrivial dependence on the ratio of two energy scales: the Zeeman energy $\Delta_z$ and the hybridization 
energy $\Delta_t$.  
Namely, we have shown that a thin film with hybridization between the top and bottom surfaces undergoes a quantum 
phase transition from a ``trivial insulator" state, in which the Hall conductivity at charge neutrality is zero 
to a "Hall insulator" state, in which the Hall conductivity is equal to $e^2/h$.
The transition happens when $\Delta_z = \Delta_t$ and is associated with a nonanalytic contribution to the Hall 
conductivity, which comes entirely from the $n = 0$ LL.
The transition is accompanied by a change in the character of the $n = 0$ LL. On the low-field side 
of the transition one of the $n = 0$ sublevels is electron-like, while the second one is hole-like. 
On the high-field side both LL are hole-like.

Finally, let us briefly discuss experimental observability of the proposed effect. 
To this end we estimate the energy scales, associated with the TI film LL. 
Taking $B \sim 10$~T, we obtain $\Delta_z \sim 1$~meV (using $g \sim 1$ for a very conservative estimate), and 
$\omega_B \sim 10$~meV. Assuming $\Delta_t \sim \Delta_z$ (which is needed to observe the phase transition we have described), 
we then have $\omega_B \gg \Delta_t, \Delta_z$.
Thus observing the ``trivial insulator" to ``Hall insulator" phase transition might require more stringent conditions than a general observation 
of the QHE in this system would require. One needs low, i.e. less than about 10~K temperatures, and 
a clean enough sample, that a $\sim 1~\textrm{meV}$ hybridization gap between the two $n=0$ LL is not washed out by disorder. 
However, recent experiments actually suggest a much larger value of the $g$-factor for the surface states in $\textrm{Bi}_2 \textrm{Se}_3$ 
and related materials,~\cite{Analytis10,Ando10} as large as $g \approx 50$ (which is not unexpected due to the strong spin-orbit interactions
in the bulk material~\cite{Kohler75}), and this makes the proposed phase transition much more easily observable.
But even the most stringent conditions, required if $g \sim1 $,  are in fact standard in semiconductor 2DEG QHE measurements and thus should be easily achievable, at least for Molecular Beam Epitaxy grown TI films.

\begin{acknowledgments}
We acknowledge useful discussions with J.~Alicea, L.~Balents, J.~Eisenstein, and O.~Motrunich. 
Financial support was provided by the NSERC of Canada and a University of Waterloo start-up grant.
AAB gratefully acknowledges hospitality of the Caltech Physics Department, where part of this work was performed. 
\end{acknowledgments}

\end{document}